\documentclass[pmlr]{jmlr}

\usepackage{longtable}

\usepackage{booktabs}
\usepackage{booktabs}

\jmlrvolume{260}
\jmlryear{2024}
\jmlrworkshop{ACML 2024}
\jmlrpages{-}
\renewcommand{\thepage}{}

\title[InfantCryNet]{InfantCryNet: A Data-driven Framework for Intelligent Analysis of Infant Cries}

 \author{\Name{Mengze Hong} \Email{mengze.hong@connect.polyu.hk}\\
 \addr Hong Kong Polytechnic University
 \AND
 \Name{Chen Jason Zhang} \Email{jason-c.zhang@polyu.edu.hk}\\
\addr Hong Kong Polytechnic University
\AND
 \Name{Lingxiao Yang} \Email{lingx.yang@polyu.edu.hk}\\
\addr Hong Kong Polytechnic University
\AND
 \Name{Yuanfeng Song} \Email{yfsong@webank.com}\\
\addr WeBank Co., Ltd. 
\AND
 \Name{Di Jiang} \Email{dijiang@webank.com}\\
\addr WeBank Co., Ltd. 
}

\editors{Vu Nguyen and Hsuan-Tien Lin}

\begin{document}

\maketitle

\begin{abstract}
Understanding the meaning of infant cries is a significant challenge for young parents in caring for their newborns. The presence of background noise and the lack of labeled data present practical challenges in developing systems that can detect crying and analyze its underlying reasons. In this paper, we present a novel data-driven framework, ``InfantCryNet,'' for accomplishing these tasks. To address the issue of data scarcity, we employ pre-trained audio models to incorporate prior knowledge into our model. We propose the use of statistical pooling and multi-head attention pooling techniques to extract features more effectively. Additionally, knowledge distillation and model quantization are applied to enhance model efficiency and reduce the model size, better supporting industrial deployment in mobile devices. Experiments on real-life datasets demonstrate the superior performance of the proposed framework, outperforming state-of-the-art baselines by 4.4\% in classification accuracy. The model compression effectively reduces the model size by 7\% without compromising performance and by up to 28\% with only an 8\% decrease in accuracy, offering practical insights for model selection and system design. 
\end{abstract}


\section{Introduction}

It is reported that more than 380,000 babies are born each day globally. For these newborns, the cry is their fundamental mode of communication with the outside world, serving as a critical indicator of their biological and psychological needs \citep{LockhartBouron2023InfantCC}. Experienced medical practitioners have developed the ability to interpret various cry sounds, enabling them to discern an infant's physical condition and, in some instances, identify potential diseases solely based on the characteristics of the cry. However, acquiring such a nuanced understanding quickly is impractical for new parents due to the lack of specific knowledge and experience \citep{jiang2023probabilistic}. 

There is a pressing need for the development of accessible, user-friendly tools that can assist parents in comprehending their newborns' cries and supporting the overall well-being of the infant \citep{LAGUNA2023107626}. However, this is challenging due to several factors.  
First, due to the presence of background noise in the living environment, effectively detecting the infant's cry is difficult \citep{9746096, 10145599}. Second, the diversity of newborn cry patterns complicates pinpointing the precise reasons for crying. Third, audio records of infant cries are typically rare, and those with annotated information are even harder to obtain, leading to data scarcity issues. Lastly, the detection and analysis of infant cries need to be done with low computing resources to enable deployment on mobile devices such as smartphones or tablets.

In this paper, we propose a novel framework named \textbf{InfantCryNet} that detects and analyzes infant cries seamlessly. Specifically, detection aims to identify whether an infant is crying in an audio clip, and analysis aims to identify the underlying needs (or reasons) of the cry. The proposed framework employs a state-of-the-art pre-trained audio model. It incorporates methods such as feature extraction, statistic pooling layers, and model compression to offer an innovative, industry-standard solution. The contributions of this work are summarized as follows:

\begin{itemize}
\item To overcome the data scarcity issue, a pre-trained model is employed to provide prior knowledge for downstream task solving, resulting in significant improvements over baseline models.
\item To effectively extract features from audio clips of varying lengths, we propose the use of statistic pooling and multi-head attention pooling methods, achieving state-of-the-art performance among global pooling techniques.
\item In an effort to enhance model efficiency and facilitate industrial applications, we experimented with the integration of model compression using knowledge distillation and model quantization, resulting in lightweight models with satisfactory performance. 
\end{itemize}

The rest of this paper is organized as follows. Section~\ref{sec:Related Work} reviews the relevant literature. Section~\ref{sec:Model} describes the model architecture and proposed techniques. 
Section~\ref{sec:experiments} reports the experimental results and discussions, and Section~\ref{sec:Conclusion} concludes with limitations and recommendations for further research.

\section{Related Work}

\label{sec:Related Work}

This study has close ties to three areas of research: audio classification, model pre-training, and model compression.

\subsection{Audio Classification with CNNs}
The task of infant cry classification has seen significant advancements in recent years, evolving from traditional machine learning techniques to deep learning approaches. Among these methods, Convolutional Neural Networks (CNNs) have gained increasing popularity due to their effectiveness in extracting local patterns. Numerous applications have been developed based on CNNs, including acoustic event detection \citep{bae2016acoustic}, music start detection \citep{schluter2014improved}, automatic speech recognition \citep{abdel2014convolutional, 10.1145/3503161.3547731, 9679032}, and causal inference \citep{du2024estimatingpeerdirectindirect, du2024causalgnnsgnndriveninstrumental}. Additionally, Graph Convolutional Networks (GCNs) have garnered interest for their ability to improve accuracy with limited labeled training data \citep{chen2024multi, ABBASKHAH2023105261}.



\subsection{Pretraining Model for Audio Tasks}
In computer vision, models commonly undergo pre-training on labeled datasets such as ImageNet \citep{deng2009imagenet}, which has a massive sample size. In natural language processing, pretraining methods based on Transformers \citep{vaswani2017attention} have been proposed, leading towards the trending Large Language Models (LLMs) \citep{Lin2024DrawandUnderstandLV}. Motivated by these advances, audio pretraining has become a popular research topic. For instance, Wav2vec \citep{schneider2019wav2vec} can be used to transfer audio signals into vector representations, which can improve the training of acoustic models. Similar to Imagenet, AudioSet \citep{gemmeke2017audio} is a massive audio dataset comprising more than 2 million audio samples, each tagged with 527 sound event labels. By pretraining on AudioSet, researchers \citep{kong2018audio, kong2020panns} have proposed a variety of deep neural networks for audio classification and achieved promising performance.

In infant cry classification, labeled datasets are scarce due to the sensitivity of data collection and the high cost of annotation by pediatricians. Extracting features from pre-trained models can effectively incorporate prior knowledge. The transfer learning technique leverages the rich feature representations learned by deep neural networks to train classifiers \citep{5288526}, thereby enhancing the performance for more specific tasks.

\subsection{Model Compression}
In the evolving landscape of machine learning techniques, there has been a notable shift in research focus from the development of "best" performing models to those that offer better efficiency and practicality for industrial applications \citep{choudhary2020comprehensive}. The advent of novel compression techniques such as knowledge distillation, network pruning, and quantization have enabled the deployment of sophisticated models in many practical use cases \citep{hinton2015distilling, zhou2017incremental}. Moreover, with the emergence of large pre-trained models, \cite{pmlr-v119-li20m} suggests the strategy of training very large models and then compressing them to obtain a relatively smaller model, which has higher accuracy than directing training a smaller model. Similar work has also shown that model compression techniques can effectively reduce the size of neural networks without significantly compromising accuracy \citep{polino2018model}, which is particularly beneficial for scenarios where computational resources are at a premium, such as in edge computing environments. 

\vspace{1ex}
Our work diverges from existing infant cry systems by addressing data scarcity with pre-trained models, enhancing CNN pooling methods for better feature extraction, and providing a lightweight, mobile-deployable solution through model compression.

\section{Framework Architecture}

\label{sec:Model}
In this section, we first introduce the feature extraction methods for infant cry audio in Section~\ref{sec:feature}. The two targeted tasks, namely crying detection and crying analysis, are introduced in Section~\ref{sec:detect} and~\ref{sec:classify}, together with the proposed pooling techniques. The model compression and compression techniques are introduced in Section~\ref{sec:accelerate}.

\subsection{Feature Extraction}
\label{sec:feature}

In audio processing, the feature extraction can be divided into time and frequency domains. Time-domain features, such as amplitude and zero-crossing, provide straightforward insights but limited information for complex tasks due to their simplicity. In contrast, frequency domain features, including MFCCs, LPCCs, and LFCCs, have been proven to achieve superior results with more discriminative features to be learned by the model \citep{Hertel2016ComparingTA}. These features can be represented by two approaches: the \textbf{waveform} that reflects the pattern of sound pressure amplitude in the time domain, and the \textbf{spectrogram} which visually depicts a signal's frequency spectrum.

\begin{figure}[t]
\centering
\includegraphics[width=\textwidth]{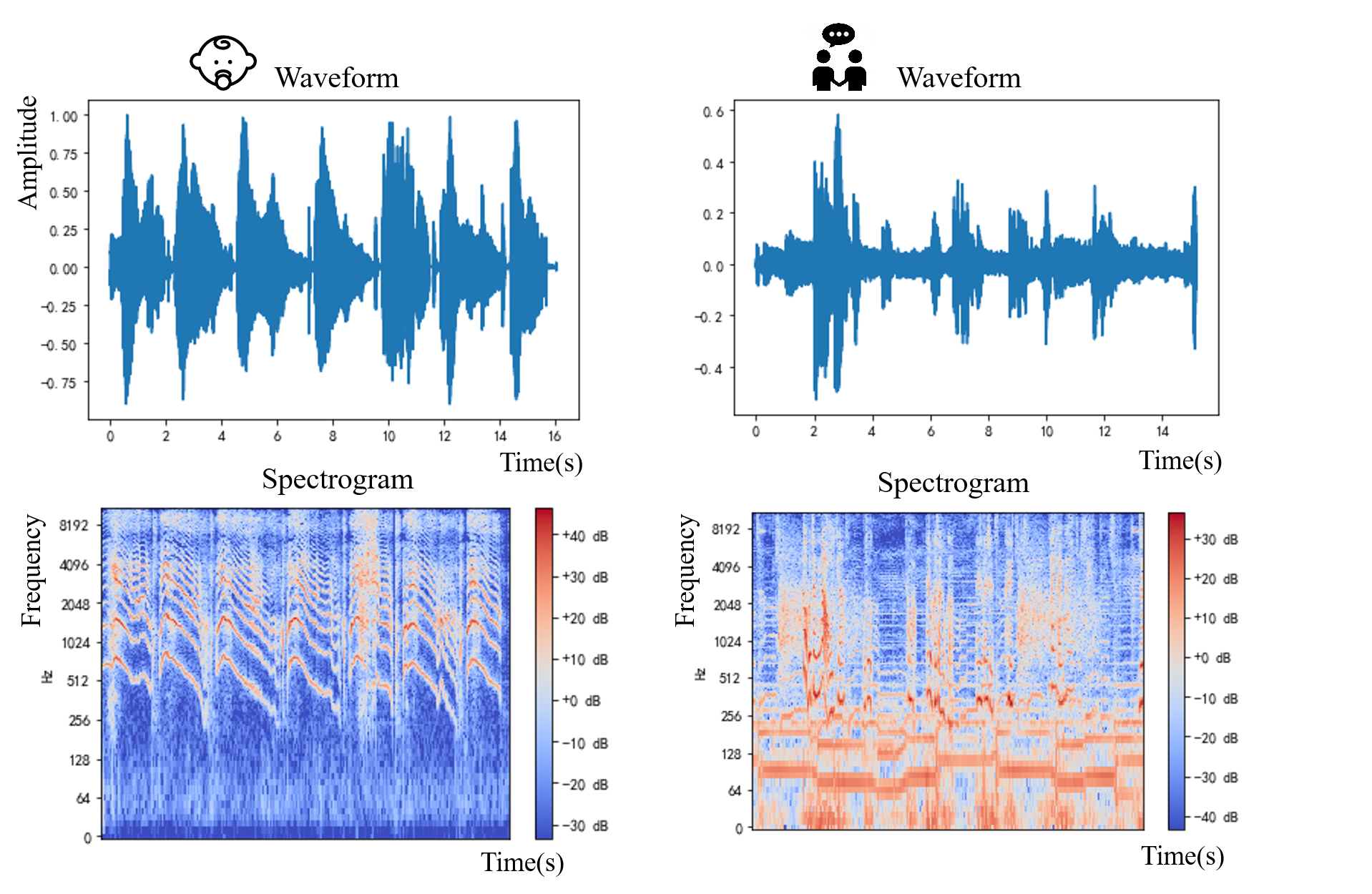}
\vspace{-2ex}
\caption{Infant cry (left) vs. adult voice (right) in waveform and spectrogram}
\label{feature}
\end{figure}

The comparison between infant cry sounds and adult voices is illustrated in Figure~\ref{feature}, using both waveform and spectrogram representations. Unlike adult speech, which tends to be more irregular and has lower amplitude, infant cries typically exhibit a more prosodic waveform and larger amplitude. This periodic nature makes infant cries particularly well-suited for combined analysis of prosodic and time-frequency domain features. The spectrogram, as a comprehensive visual representation, effectively displays both acoustic and prosodic characteristics, providing clear insights into how the signal's frequency content varies over time and highlighting the unique properties of infant cries.

With this feature representation, the problem of classifying audio has been transformed into an image classification problem, motivating the following discussions on constructing convolutional neural networks. Determining the right network architecture is crucial for developing the infant cry system. Thus, we present two distinct model architecture options in the subsequent sections, specifically focusing on the tasks of infant cry detection and classification. These proposed architectures aim to effectively capture relevant information and address the aforementioned challenges in the infant cry system.

\begin{figure}[th!]
\centering
\includegraphics[width=\textwidth]{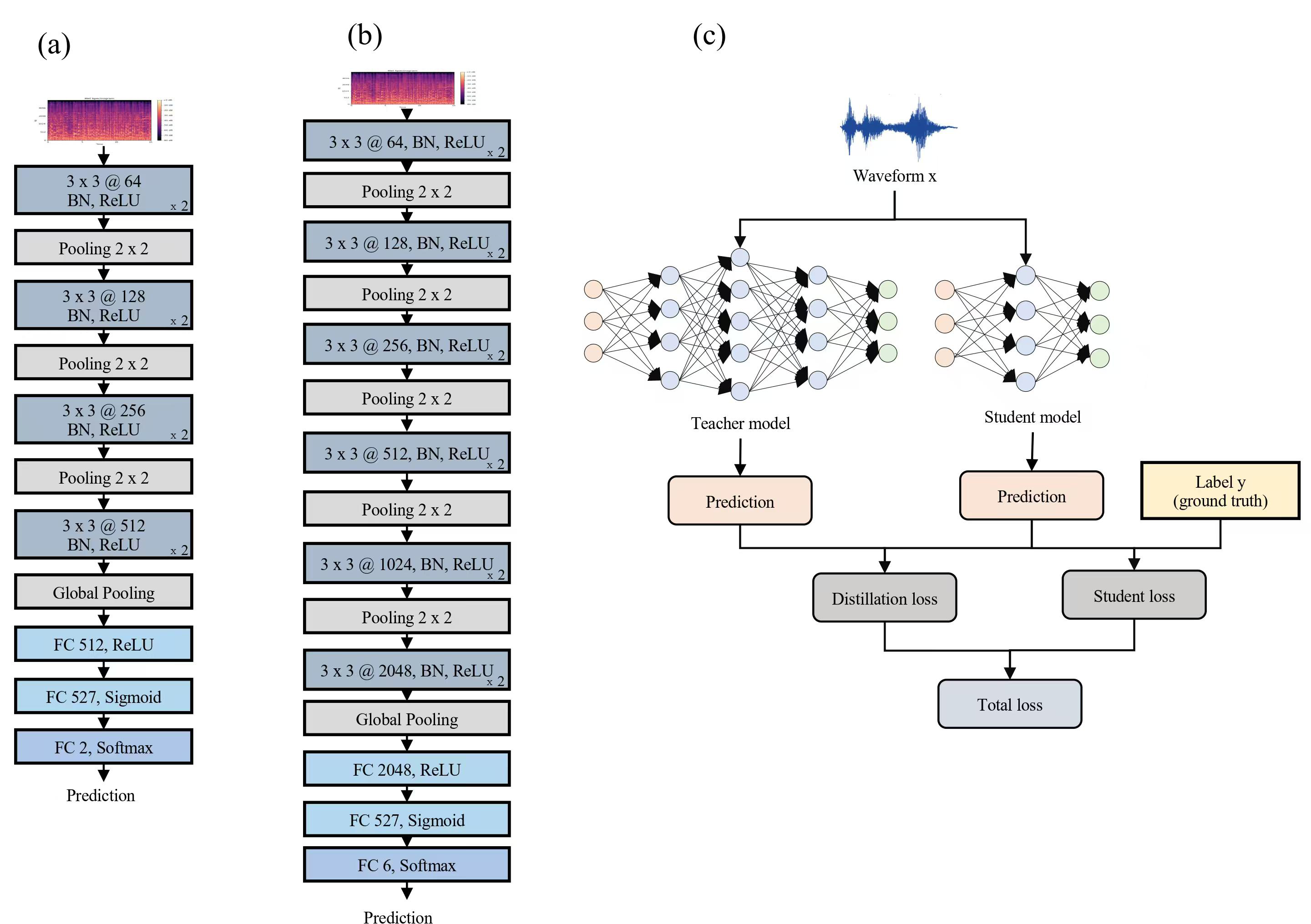}
\vspace{-2ex}
\caption{Model architect: (a) CNN10 for infant cry detection, (b) CNN14 for infant cry classification, (c) Knowledge Distillation for model compression.}
\label{model}
\end{figure}

\subsection{Infant Cry Detection}
\label{sec:detect}

For infant cry detection, as shown in Figure~\ref{model}(a), we utilize the 10-layer CNNs (CNN-10) with four convolutional blocks, each consisting of two $3\times3$ kernel-sized convolutional layers, separated by batch normalization to improve training efficiency and stability. The ReLU activation function is utilized for this purpose. After each block, the spatial dimensions of the feature maps are reduced via a $2\times2$ average pooling operation. The final feature maps are summarized into a fixed-length vector through another pooling operation and fed into a softmax activation function to generate class probabilities for the binary classification of whether an audio clip contains an infant cry.

\subsection{Infant Cry Analysis}
\label{sec:classify}

Based on the available dataset, the infant cries can be classified into six distinct reasons: \textit{awake}, \textit{hug}, \textit{sleepy}, \textit{uncomfortable}, \textit{diaper}, and \textit{hungry}. Here, we utilize the 14-layer CNNs (CNN-14), as shown in Figure~\ref{model}(b), which has two more convolutional blocks in comparison with CNN-10  for enhanced feature extraction and improved classification accuracy in identifying the distinct reasons behind infant cries. 

Suppose the segment level embedding is $\boldsymbol{H}=\{\boldsymbol{h}_{1}, \boldsymbol{h}_{2}, ..., \boldsymbol{h}_{N}\}$. The common pooling methods include max pooling and average pooling. Max pooling selects the maximum value in each patch, with the limitation that only one instance represents the whole audio while other instances are ignored. This can be described as follows:
\begin{align}
    {h}_{max} = max(\boldsymbol{h}_{1}, \boldsymbol{h}_{2}, ..., \boldsymbol{h}_{N})
\end{align}
Average pooling calculates the average value of each patch, which considers all instances instead of the maximum. However, since each instance contributes equally, it fails to reflect the variance within the patch: 
\begin{align}
    {h}_{avg}=\frac{\sum_{i=1}^N\boldsymbol{h}_{i}}{N}
\end{align}
In order to combine the advantages of max pooling and average pooling, the pre-trained audio neural networks (PANNs) calculated a fixed-length vector by adding the averaged and maximized vectors \citep{kong2020panns}.
However, the difference between instances is still ignored, and the method is formulated as follows:
\begin{align}
    {h}_{add}=max(\boldsymbol{h}_{1}, \boldsymbol{h}_{2}, ..., \boldsymbol{h}_{N})+\frac{\sum_{i=1}^N\boldsymbol{h}_{i}}{N}
\end{align}

Recognizing the limitations of existing methods, we propose two kinds of global pooling methods: statistic pooling and multi-head attention pooling. 
\begin{itemize}

\item \textbf{Statistic pooling} calculates the mean and variance, then uses a linear layer to reduce two dimensions to one. Both the average and variability of values are leveraged to represent the feature. This can be summarized as:
\begin{align}
    &s^2=\frac{\sum_{i=1}^N(\boldsymbol{h}_{i}-{h}_{avg})^2}{N}
\end{align}
\begin{align}
{h}_{stat}=&fc({h}_{avg},s^2)
\end{align}
where $fc(*)$ represents linear layer and ${h}_{avg}$ denotes the average pooling.

\item \textbf{Multi-head attention pooling} determines the attention distribution of each instance, allowing instances to be aggregated based on their significance rather than equally contributing to the feature representation, as shown below:
\begin{align}
    {h}_{attn}=\frac{\sum_{i=1}^N({\alpha}_{i}\boldsymbol{h}_{i})}{\sum_{i=1}^N({\alpha}_{i})}
\end{align}
where $\alpha=Wh$, $h$ denotes the output from the previous layer, and $W\in \mathbb{R}^{N\times N}$ is the learned attention parameter.
\end{itemize}

\subsection{Model Compression}
\label{sec:accelerate}

For challenging tasks, models with complex network architectures typically perform better. However, the computation for such networks can be slow, posing the challenge of achieving good performance with a lightweight network. This paper addresses this problem by employing knowledge distillation \citep{hinton2015distilling}, which transfers knowledge from a complex "teacher" model to a simpler "student" model without sacrificing accuracy. It also considers model quantization technique \citep{jacob2018quantization}, which performs some or all operations on tensors using integers instead of floating-point values.
 
For knowledge distillation, we utilize ResNet22 \citep{he2016deep} as the student model and CNN14 as the teacher model, with the distillation process illustrated in Figure~\ref{model}(c). The loss function for the student model is carefully crafted to address two main discrepancies. First, it measures the divergence between the predictions made by the student model and the actual ground truth data, ensuring that the student learns to approximate the correct outputs effectively. Second, it accounts for the variance between the softened predictions of the student model and the softened labels provided by the teacher model. These softened labels are generated by applying a temperature scaling factor to the teacher's logits, which allows the student model to capture more nuanced information about the output distribution.
 
On the other hand, we implement dynamic quantization, which involves converting the weights and activation of the model from floating-point to integer values with scale factors determined dynamically at runtime, leading to a smaller model size and faster inference.

\section{Experiments}
\label{sec:experiments}

In this section, we outline the experiments conducted to evaluate the performance of the proposed methods. We detail the datasets used, the baselines for comparison, and the implementation specifics. Additionally, we conduct several ablation studies to analyze the effectiveness of different components, suggesting best practices for implementing the system.

\subsection{Experimental Setup}
\subsubsection{Dataset}

\begin{table}[t]
\centering
\begin{tabular}{ccc}
    \toprule
    \textbf{Task} & \textbf{Training} & \textbf{Testing} \\
    \midrule
    infant cry detection & 6,000 & 600 \\
    infant cry classification & 750 & 85 \\
    \bottomrule
\end{tabular}
\caption{Overview of dataset}
\label{table:dataset}
\end{table}

In the task of infant cry detection, a dataset consisting of 6,600 audio clips is used, with 6,000 clips for training and 600 clips for validation (see Table \ref{table:dataset}). Each audio clip has a 15-second duration and a sampling rate of 16,000 Hz. Similarly, the dataset for the infant cry classification task contains 835 audio clips, each labeled by one of six reasons for crying (i.e., awake, hug, sleepy, uncomfortable, diaper, and hungry).

\subsubsection{Baselines}
The performance of the proposed model is compared with various baselines and reported in terms of classification accuracy. The baseline models include:

\begin{itemize}
\item \textbf{CNN10} The 10-layer CNN, which features only four convolutional layers, is a simpler architecture compared to CNN14, which has six convolutional layers. 

\item \textbf{Resnet22} Each block in the ResNet comprises two convolutional layers with $3 \times3$ kernel sizes and a shortcut connection between convolutional layers. Here, a 22-layer deep ResNet (Resnet22) with eight basic blocks is considered.

\item \textbf{Wavegram-Logmel-CNN} In comparison to one-dimensional CNN that cannot capture frequency information, \cite{kong2020panns} proposed Wavegram-Logmel-CNN, which combined wavegram (time-domain) and Log-Mel spectrogram (frequency-domain) as input to a CNN.
\end{itemize}

\subsubsection{Implementation Details}

The infant cry detection task involves identifying the presence of crying activity in an audio clip. To accomplish this, we first extracted the Log-Mel spectrogram, which serves as the input features for our model, and then fine-tuned a 10-layer CNN from PANNs for this purpose. The infant cry classification involves categorizing audio clips according to the reasons for crying. Similarly, we begin by extracting Log-Mel spectrograms, followed by fine-tuning a 14-layer CNN from PANNs to build an effective classification system. The Adam optimizer is used for both models, and the training was conducted on eight Nvidia Tesla V100 32GB GPUs with a batch size of 256.

\label{subsec:result}

\begin{table}[t]
\centering
\begin{tabular}{lc}
\toprule
\textbf{Model} & \textbf{Accuracy}\\
\midrule
CNN10 & 99.8\%  \\
CNN14  & 99.8\%  \\
\bottomrule
\end{tabular}
\caption{Performance of infant cry detection}
\label{table:detect}
\end{table}

\subsection{Accuracy Analysis}

For infant cry detection, the results of our proposed method and a baseline model are shown in Table~\ref{table:detect}, where both models exhibit similar accuracy. This can be attributed to the fact that crying sounds are relatively distinct and easy to detect compared to other sounds, enabling even a simple network to perform effectively. The high amplitude observed in Figure~\ref{feature} contributes to the mitigation of the effect of background noise in the audio. Hence, it is advisable to choose computationally efficient models that can still achieve good performance in detecting infant cries.

\begin{table}[t]
\centering
\begin{tabular}{lc}
\toprule
\textbf{Model}& \textbf{Accuracy} \\
\midrule
Wavegram-Logmel-CNN with pretraining & 70.33\%  \\
Resnet22 with pretraining & 53.85\% \\
CNN10 with pretraining & 52.75\% \\
CNN14  & 64.84\%  \\
\textbf{CNN14 with pretraining} & \textbf{74.73\%}  \\
\bottomrule
\end{tabular}
\caption{Performance of infant cry classification with different network architectures}
\label{table:classify}
\end{table}

For the classification task, the comparison between our method (i.e., CNN14 with pretraining) and various baselines is presented in Table \ref{table:classify}. The proposed model outperforms all baselines, achieving an encouraging 4.4\% improvement in accuracy over the best-performing method. This demonstrates the effectiveness of the proposed network architecture for solving the classification problem, and these results support the hypothesis that a pre-trained model can significantly improve performance compared to those without pretraining.

\begin{table}[t]
\centering
\begin{tabular}{lc}
\toprule
\textbf{Pooling Method} & \textbf{Accuracy}\\
\midrule
Max pooling & 62.64\%  \\
Average pooling & 70.33\%  \\
Max pooling + average pooling & 71.42\%  \\
\textbf{Statistic pooling} (ours) & \textbf{74.73\%}  \\
Multi-head attention pooling (ours)  & 72.53\%  \\
\bottomrule
\end{tabular}
\caption{Performance of infant cry classification with different pooling methods}
\label{table:pooling}
\end{table}

\subsection{Ablation Study}

By focusing on the accuracy comparison of the same model (i.e., CNN14 with pretraining) under different pooling methods (see Table \ref{table:pooling}), our findings revealed that average pooling consistently outperformed max pooling among the baseline methods, underscoring the advantage of incorporating all instances in the calculation. While the combination of max pooling and average pooling provided only a small improvement, our proposed methods demonstrated superior performance. Notably, statistic pooling achieved the highest accuracy, which can be attributed to the periodic nature of infant cries, where varying attention to instances is not always necessary. Attention pooling ranked as the second-best option. Overall, our proposed methods outperformed the baselines, achieving significant improvements in accuracy over the best-performing baseline method.

\begin{table}[t]
\centering
\begin{tabular}{lcc}
\toprule
\textbf{Model}& \textbf{Accuracy}& \textbf{Model Size}\\
\midrule
CNN14 (teacher) & 73.63\% & 81M\\
Resnet22 (student) & 53.85\% & 64M\\
Knowledge distillation & 68.23\% & 64M\\
Model quantization & 73.63\% & 75M\\
Model quantization + distillation  & 67.42\%  & 58M\\
\bottomrule
\end{tabular}
\caption{Effectiveness of model compression in infant cry classification}
\label{table:model compression}
\end{table}

\subsection{Model Compression}

The results in Table \ref{table:model compression} depict the effectiveness of the model compression techniques. By applying model quantization, the model size is reduced by 7\% without any loss of accuracy. Knowledge distillation reduces the model size by 21\% but also lowers the accuracy by 7\%. By combining knowledge distillation with model quantization, the model size is reduced by 28\%, and the accuracy is only reduced by 8\%.

Based on the observations, we conclude that model compression techniques can effectively reduce model size with a manageable compromise on accuracy, depending on the method chosen. Model quantization is ideal if the model size requirement is not strict, as it reduces size without losing accuracy. However, for significant size reduction, combining model quantization with knowledge distillation is effective, though it may slightly decrease accuracy. This paper suggests that the choice of compression technique should align closely with the application's specific needs. For the infant cry system, given the increasing capability of mobile devices, model quantization is recommended to retain optimal accuracy.

\section{Conclusion}
\label{sec:Conclusion}

This paper presents ``InfantCryNet,'' an innovative framework designed to detect and comprehend the meanings behind infant cries, addressing the challenges of background noise and limited labeled data. By utilizing pre-trained audio models, along with statistical and multi-head attention pooling techniques, we enhanced feature extraction capabilities and achieved significant improvement in classification accuracy. To facilitate the deployment on mobile devices, our approach incorporates knowledge distillation and model quantization to compress model size, offering practical solutions to real-world applications.


While the experiments present promising results, this study extensively focused on comparing neural network solutions with varying complexity levels. Although traditional machine learning methods generally lag in accuracy, their lightweight nature and training efficiency suggest the potential for exploring hybrid models, such as CNN-SVM and GMM-CNN, for developing more efficient infant cry classification systems. Furthermore, the challenge of data scarcity can be mitigated by adopting a federated learning approach, which enables the collection of more training data while protecting end-user privacy \citep{10.1145/3447687, 10.1007/978-3-030-59419-0_54, 10.1145/3357384.3357909}.

\bibliography{acml24}

\end{document}